\documentclass[sigconf]{acmart}

\copyrightyear{2024}
\acmYear{2024}
\setcopyright{rightsretained}
\acmConference[ICSE '24]{2024 IEEE/ACM 46th International Conference on Software Engineering}{April 14--20, 2024}{Lisbon, Portugal}
\acmBooktitle{2024 IEEE/ACM 46th International Conference on Software Engineering (ICSE '24), April 14--20, 2024, Lisbon, Portugal}\acmDOI{10.1145/3597503.3639187}
\acmISBN{979-8-4007-0217-4/24/04}

\AtBeginDocument{%
  \providecommand\BibTeX{{%
    \normalfont B\kern-0.5em{\scshape i\kern-0.25em b}\kern-0.8em\TeX}}}

\newcommand{\tool}{GILT\xspace}

\newcommand{\fulltool}{(Generation-based Information-support with LLM Technology)\xspace}

\usepackage{amsmath,amsfonts}
\usepackage{algorithmic}
\usepackage{textcomp}
\usepackage{xcolor}
\usepackage{graphicx}
\usepackage{hyperref}
\usepackage{cleveref}
\usepackage{booktabs}
\usepackage{subfig}
\usepackage{xspace}
\usepackage{multirow}
\usepackage{makecell}
\usepackage{url}
\usepackage{enumitem}
\usepackage{adjustbox}
\usepackage{balance}
\usepackage{multicol}
\usepackage{graphicx}
\usepackage{url}
\usepackage{listings}
\usepackage{parcolumns}
\usepackage[most]{tcolorbox}
\usepackage[normalem]{ulem}

\usepackage{tabularx} 
\usepackage{array}
\usepackage{tabularray}
\newcolumntype{L}[1]{>{\raggedright\let\newline\\\arraybackslash\hspace{0pt}}m{#1}}
\newcolumntype{C}[1]{>{\centering\let\newline\\\arraybackslash\hspace{0pt}}m{#1}}
\newcolumntype{R}[1]{>{\raggedleft\let\newline\\\arraybackslash\hspace{0pt}}m{#1}}

\usepackage{rotating}

\usepackage{siunitx}

\usepackage{color, colortbl}
\definecolor{Gray}{gray}{0.9}

\usepackage{ifthen}
\newboolean{showcomments}
\setboolean{showcomments}{true} 
\ifthenelse{\boolean{showcomments}}
  {\newcommand{\nb}[2]{
    \fcolorbox{gray}{yellow}{\bfseries\sffamily\scriptsize#1}
    {\sf\small$\blacktriangleright$\textit{#2}$\blacktriangleleft$}
  }
  
  }
  {\newcommand{\nb}[2]{}
  
  }

\newtcbtheorem{Summary}{\bfseries Summary}{enhanced,drop shadow={black!50!white},
  coltitle=black,
  top=0.15in,
  attach boxed title to top left=
  {xshift=1.5em,yshift=-\tcboxedtitleheight/2},
  boxed title style={size=small,colback=white}
}{summary}

\newtcolorbox[auto counter]{summary}[1][]{title={\bfseries Summary~#1},enhanced,drop shadow={black!50!white},
  coltitle=black,
  top=0.1in,
  attach boxed title to top left=
  {xshift=1.5em,yshift=-\tcboxedtitleheight/2},
  boxed title style={size=small,colback=white},}

\newcommand{\eg}{e.g.,\xspace}
\newcommand{\ie}{i.e.,\xspace}
\newcommand{\vs}{vs.\xspace}

\newcommand{\etal}{\emph{et al.}\xspace}

\usepackage{grffile}
\begingroup\catcode`\%=12
\def\x{\def\strangepath{
\expandafter\endgroup\x

\makeatletter
\def\thickhline{%
  \noalign{\ifnum0=`}\fi\hrule \@height \thickarrayrulewidth \futurelet
  \reserved@a\@xthickhline}
\def\@xthickhline{\ifx\reserved@a\thickhline
              \vskip\doublerulesep
              \vskip-\thickarrayrulewidth
             \fi
      \ifnum0=`{\fi}}
\makeatother

\newlength{\thickarrayrulewidth}
\setlength{\thickarrayrulewidth}{2\arrayrulewidth}

\newcounter{RQCounter}
\newcounter{HCounter}
\newcounter{RSCounter}
\newcommand{\RQ}[2]{%
\refstepcounter{RQCounter} \label{#1}
 	\vspace{0.02in}
    \noindent \textbf{RQ\arabic{RQCounter}.~#2
	\vspace{0.05in}
    }
}

\usepackage{framed}
\newcommand{\RS}[2]{%
\refstepcounter{RSCounter} \label{#1}
\begin{framed}%
\filbreak
\noindent \textbf{Result \arabic{RSCounter}}:~{\emph {#2}}%
\end{framed}
}

\newcommand{\Hyp}[2]{%
\refstepcounter{HCounter} \label{#1}
	\vspace{0.015in}
	\noindent 
	\textbf{H}$_{\arabic{HCounter}}$.~\emph{#2}
	\vspace{0.025in}
}
\newcommand{\hr}[1]{\textbf{H}$_{\ref{#1}}$}
\newcommand{\RQref}[1]{\textbf{RQ\ref{#1}}}
\newcommand{\mysec}[1]{\vspace{0.1cm} \noindent \textbf{#1.}}
\newcommand{\mysubsec}[1]{\vspace{0.05cm} \emph{#1.}}
\newcommand{\variable}[1]{{\small\textbf{\fontfamily{ppl}\selectfont#1}}}

\newcommand{\revision}[1]{{#1}\xspace}

\newcommand{\edit}[2]{{#2} \xspace}

\newcommand{\remove}[1]{{}}

\newcommand{\cameraready}[1]{{#1}\xspace}

\usepackage[normalem]{ulem}

\usepackage{tikz}
\newcommand*\circled[1]{\tikz[baseline=(char.base)]{
            \node[shape=circle,draw,inner sep=0.2pt] (char) {#1};}}
\newcommand*\bcircled[1]{\tikz[baseline=(char.base)]{
            \node[shape=circle,fill,inner sep=0.2pt] (char) {\textcolor{white}{#1}};}}

\definecolor{ABlue}{HTML}{127bca}
\definecolor{LHScolor}{HTML}{555555}

\usepackage[scaled]{helvet}
\usepackage[most]{tcolorbox}

\newcommand{\droptextshadow}[2]{%
    \tikz[baseline,outer sep=0pt, inner sep=0pt]{
    \node[#1!40!black] at (0,-0.1ex) {#2};
    \node[white] at (0,0) {#2};
}%
}

\newcommand{\DOIbox}[1]{
\tcbsidebyside[
        bicolor,
        sidebyside,
        fontupper=\footnotesize\ttfamily\bfseries,
        fontlower=\footnotesize\sffamily\mdseries,
        nobeforeafter,
        shrink tight,
        extrude bottom by=1mm,
        sidebyside adapt=both,
        sidebyside gap=5pt,
        top=2pt,left=3pt,right=3pt,bottom=2pt,
        boxrule=0pt,
        rounded corners,
        coltext=white,
        colback=LHScolor,
        colbacklower=ABlue,
]{%
DOI
}{%
\href{http://dx.doi.org/#1}{#1}
}%
}

\begin{document}

\title{Using an LLM to Help With Code Understanding}

\begin{abstract}
Understanding code is challenging, especially when working in new and complex development environments. Code comments and documentation can help, but are typically scarce or hard to navigate. Large language models (LLMs) are revolutionizing the process of writing code. Can they do the same for helping understand it?
In this study, we provide a first investigation of an LLM-based conversational UI built directly in the IDE that is geared towards code understanding. Our IDE plugin queries OpenAI's GPT-3.5-turbo model with four high-level requests \textit{without} the user having to write explicit prompts: to explain a highlighted section of code, provide details of API calls used in the code, explain key domain-specific terms, and provide usage examples for an API. The plugin also allows for open-ended prompts, which are automatically contextualized to the LLM with the program being edited.
We evaluate this system in a user study with 32 participants, which confirms that using our plugin can aid task completion more than web search. We additionally provide a thorough analysis of the ways developers use, and perceive the usefulness of, our system, among others finding that the usage and benefits differ between students and professionals.
We conclude that in-IDE prompt-less interaction with LLMs is a promising future direction for tool builders.
\end{abstract}

\author{Daye Nam}
\affiliation{%
  \institution{Carnegie Mellon University}
  \country{U.S.A.}
}
\email{dayen@cs.cmu.edu}

\author{Andrew Macvean}
\affiliation{%
  \institution{Google, Inc.}
  \country{U.S.A.}
}
\email{amacvean@google.com}

\author{Vincent Hellendoorn}
\affiliation{%
  \institution{Carnegie Mellon University}
  \country{U.S.A.}
}
\email{vhellend@andrew.cmu.edu}

\author{Bogdan Vasilescu}
\affiliation{%
  \institution{Carnegie Mellon University}
  \country{U.S.A.}
}
\email{vasilescu@cmu.edu}

\author{Brad Myers}
\affiliation{%
  \institution{Carnegie Mellon University}
  \country{U.S.A.}
}
\email{bam@cs.cmu.edu}
\maketitle

\section{Introduction}
\label{sec:intro}
\revision{
Building and maintaining software systems requires a deep understanding of a codebase. Consequently, developers spend a significant amount of time searching and foraging for the information they need and organizing and digesting the information they find~\cite{Ko.2007, Piorkowski.2016, Ko.2006, Meyer.2017,latoza2006maintaining, Maalej.2014}.
Understanding code, however, is a challenging task; developers need to assimilate a large amount of information about the semantics of the code, the intricacies of the APIs used, and the relevant domain-specific concepts.
Such information is often scattered across multiple sources, making it challenging for developers, especially novices or those working with unfamiliar APIs, to locate what they need. 
Furthermore, much of the relevant information is inadequately documented or spread across different formats and mediums, where it often becomes outdated. 
}

\revision{With the growing popularity of large language model (LLM) based code generation tools~\cite{chatgpt,copilot,tabnine}, the need for information support for code understanding is arguably growing even higher.
These tools can generate code automatically, even for developers with limited coding skills or domain knowledge. This convenience comes at a cost, however -- developers may receive code they don't understand~\cite{Ziegler.2022, imai2022github}. 
Indeed, early research on LLM code generation tools has found that developers have a harder time debugging code generated by the LLM and easily get frustrated~\cite{liang2023understanding, Vaithilingam.2022}. 
}

\revision{Fortunately, LLMs also provide an opportunity in this space, namely by offering on-demand \textit{generation-based information support} for developers faced with unfamiliar code. Compared to general web search queries~\cite{xia2017developers}, LLM prompts can allow developers to provide more context, which can enable them to receive information that more precisely aligns with their specific needs, potentially reducing the time spent on sifting through the information obtained from the web to suit their particular requirements. Developers have indeed taken to web-hosted conversational LLM tools, such as ChatGPT, for programming support en masse, but this setup requires them to both context switch and copy the relevant context from their IDEs into the chat system for support.}


\revision{To explore the potential for generation-based information support directly in the developer's programming environment, we developed a prototype in-IDE LLM information support tool, \tool \fulltool.
\tool is capable of generating on-demand information while considering the user's local code context, which we incorporate into the prompts provided to the LLM behind the scenes.
This way, we also introduce a novel interaction method with the LLM, \textit{prompt-less interaction}. 
This option aims to alleviate the cognitive load associated with writing prompts, particularly for developers who possess limited domain or programming knowledge.}

As there is still little knowledge about how to best use an LLM for information support (as opposed to just code generation), we evaluate the effectiveness of our prototype tool in an exploratory user study with 32 participants tasked with comprehending and extending unfamiliar code that involves new domain concepts and Python APIs for data visualization and 3D rendering -- a challenging task. 
Our study quantitatively compares task completion rates and measures of code understanding between two conditions -- using the LLM-backed assistant in-IDE versus directly searching the web in a browser -- and qualitatively investigates how participants used the tools and their overall satisfaction with this new interaction mode.
Concretely, we answer three research questions:
\begin{itemize}
    \item RQ1: To what extent does \tool affect developers' understanding, task completion time, and task completion rates when faced with unfamiliar code?
    \item RQ2: How do developers interact with \tool,  \revision{and to what extent does that differ between the participants}?
    \item RQ3: How do developers perceive the usefulness of \tool?
\end{itemize}

Our results confirm that there are statistically significant gains in task completion rate when using \tool, compared to a web search, showing the utility of generation-based information support.
However, we did not find the utility gains in terms of time and understanding level, leaving room for further improvement.
We also discovered that the degree of the benefit varies between students and professionals, and investigated potential reasons behind this.



\begin{figure*}
    \centering
    \includegraphics[width=0.9\linewidth]{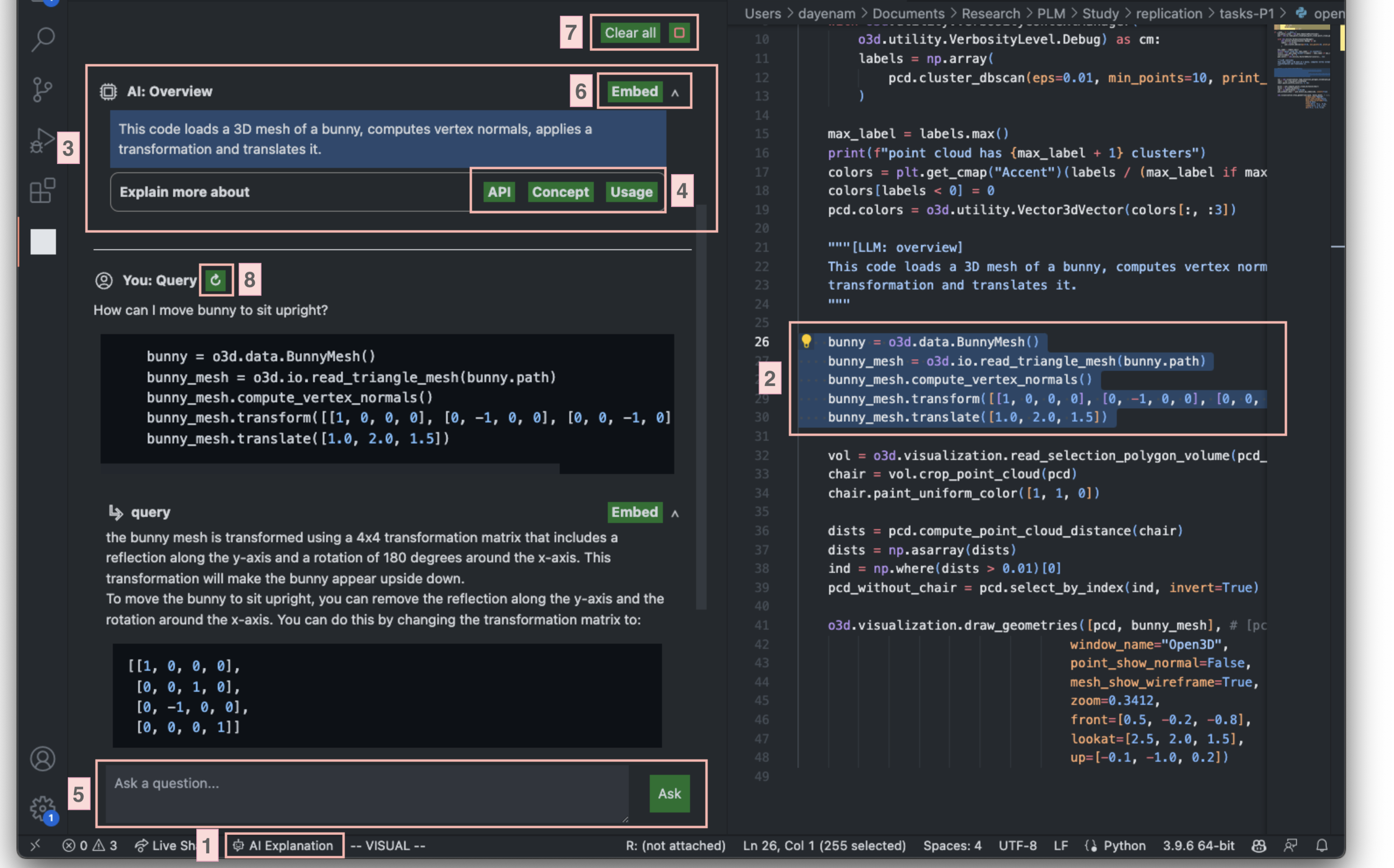}
    \caption{Overview of our prototype. (1) A trigger button; (2) code used as context when prompting LLM; (3) code summary (no-prompt trigger); (4) buttons for further details; (5) an input box for user prompts; (6) options to embed information to code (Embed) and a hide/view button; (7) options to clear the panel (Clear all) and an abort LLM button; (8) a refresh button.}
    \label{fig:tool}
\end{figure*}

\section{Related Work}
\label{sec:literature}

\subsection{\revision{Studies on Developers Information Seeking}}

In every phase of modern software engineering, developers need to work with unfamiliar code and domains, and how well they learn such code influences their productivity significantly.
Therefore, researchers have studied to understand how developers learn and comprehend unfamiliar code and domains~\cite{Nam.2023b, Ko.2007, Roehm.2015, Li.2013}, especially how they search for and acquire information, as developers need a variety of kinds of knowledge~\cite{maalej2013patterns,sillito2008asking,Thayer.2021,DBLP:conf/chi/ZhangHKG20}.
Particularly, lots of research was done on the information seeking strategies of developers, mostly in general software maintenance~\cite{Ko.2007, lawrance2010programmers, freund2015contextualizing} or web search settings~\cite{brandt2009two, rao2020analyzing}.
Researchers have also studied challenges developers face~\cite{duala2010information, Ko.2011, Rahman.2018, xia2017developers, Wang.2020, Horvath.2019}, including difficulties in effective search-query writing and information foraging, and built tools to support developers overcome such challenges~\cite{Nam.2023a, Barnaby.2020, Liu.2022}.
In this work, we explore a way of supporting developers' information needs with generation-based information support using LLMs.

Other efforts were made to understand developers' information seeking within software documentation, 
which is the main source of information for developers when they learn to use new APIs or libraries. 
Researchers cataloged problems developers face when using documentation~\cite{chen2009empirical, robillard2009makes, robillard2011field}
and identified types of knowledge developers report to be critical~\cite{robillard2009makes, robillard2011field, uddin2015api, Meng.2017},
which were taken into account when we designed \tool for our study.

\subsection{\revision{Studies on LLM-based Developer Tools}}

The potential and applicability of LLM-based AI programming tools have been actively studied by many researchers.
Numerous empirical studies~\cite{DBLP:conf/ace/Finnie-AnsleyDB22, Leinonen.2023, Hellas.2023, Sarsa.2022} evaluated the quality of code or explanations generated by LLMs, to test the feasibility of applying LLM into development tools~\cite{Ziegler.2022, Tian.2023, Jiawei.2023} and to Computer Science education~\cite{Hellas.2023, Sarsa.2022, Amoozadeh.2023}.
Several studies have also compared LLM-generated code and explanations with those authored by humans without LLM assistance~\cite{Leinonen.2023, Hellas.2023, Perry.2022}, demonstrating that LLMs can offer reasonably good help for developers or students with caution. 


Fewer studies have specifically explored the \textit{usefulness} of LLM-based programming tools~\cite{Ziegler.2022, liang2023understanding, Vaithilingam.2022, Barke.2022, Kazemitabaar.2023, Xu.2022, Mozannar.2022, Sarkar.2022, Ziegler.2022, Ross.2023} with actual users or their usage data, and many of these studies have focused on code generation tools like CoPilot~\cite{copilot}. 
For instance, Ziegler et al.\cite{Ziegler.2022} analyzed telemetry data and survey responses to understand developers' perceived productivity with GitHub Copilot, revealing that over one-fifth of suggestions were accepted by actual developers. 
Several human studies were also conducted. 
Vaithilingam~\etal~\cite{Vaithilingam.2022} compared the user experience of GitHub Copilot to traditional autocomplete in a user study and found that participants more frequently failed to complete tasks with Copilot, although there was no significant effect on task completion time. 
Barke~\cite{Barke.2022} investicated further with a grounded theory analysis to understand \textit{how} programmers interact with code-generating models, using Github Copilot as an example. 
They identified two primary modes of interaction, acceleration or exploration, where Copilot is used to speed up code authoring in small logical units or as a planning assistant to suggest structure or API calls.  

Although these studies have increased our understanding of the usefulness and usability of AI programming assistants in general, and some of the insights apply to information support, they do not show the opportunities and challenges of LLM-based tools as information support tools, with a few following exceptions~\cite{Ross.2023, MacNeil.20222dd}.
MacNeil et al.~\cite{MacNeil.20222dd} examined the advantages of integrating code explanations generated by LLMs into an interactive e-book focused on web software development, with a user study with sophomores.
They found students tend to find LLM-generated explanations to be useful, which is promising, but the study was focused on providing one-directional support in an introductory e-book which is different from user-oriented need-based information support.
The Programmer's assistant~\cite{Ross.2023} is the closest to our work.
The authors integrated a conversational programming assistant into an IDE to explore other types of assistance beyond code completion. 
They collected quantitative and qualitative feedback from a human 
study with 42 participants from diverse backgrounds and found that the \textit{perceived} utility of the conversational programming assistance was high.
In our work, we focus on the utility of LLM-based tools to satisfy information needs for \textit{code understanding}, and take a step forward to test the actual utility of an LLM-integrated programming tool by assessing performance measures such as completion rates, time, and participants' code understanding levels.

\vspace{-0.15cm}
\section{The \tool Prototype Tool}
\label{sec:prototype}

We iteratively designed \tool to explore different modes of interaction with an LLM for information support. \tool is a plugin for the VS Code IDE (Figure~\ref{fig:tool}) that considers user context (the code selected by the user) when querying an LLM for several information support applications.

\subsection{Interacting with \tool}

There are two ways to interact with the plugin. First, users can select parts of their code (Figure~\ref{fig:tool}-\circled{2}) and trigger the tool by clicking on ``AI Explanation'' on the bottom bar (Figure~\ref{fig:tool}-\circled{1}), or using ``alt/option + a'' as a shortcut, to receive a summary description of the highlighted code ({\small\texttt{Overview}}). They can then explore further by clicking on buttons (Figure~\ref{fig:tool}-\circled{4}) for API ({\small\texttt{API}}), domain-specific concepts ({\small\texttt{Concept}}), and usage examples ({\small\texttt{Usage}}), which provide more detailed explanations with preset prompts. The {\small\texttt{API}} button offers detailed explanations about the API calls used in the code, the {\small\texttt{Concept}} button provides domain-specific concepts that might be needed to understand the highlighted code fully, and the {\small\texttt{Usage}} button offers a code example involving API calls used in the highlighted code.

Users can also ask a specific question directly to the LLM via the input box (Figure~\ref{fig:tool}-\circled{5}). If no code is selected, the entire source code is used as context ({\small\texttt{Prompt}}); alternatively, the relevant code highlighted by the user is used ({\small\texttt{Prompt-context}}).
The model will then answer the question with that code as context. 
\tool also allows users to probe the LLM by supporting conversational interaction ({\small\texttt{Prompt-followup}}). When previous LLM-generated responses exist, if a user does not highlight any lines from the code, the LLM generates a response with the previous conversion as context. Users can also reset the context by triggering the tool with code highlighted, or with the Clear all button.

\subsection{Our Design Process and Decisions}


\mysec{Focus on understanding}
We intentionally did not integrate a code generation feature in the prototype as we wanted to focus on how developers \textit{understand} code.

\mysec{In-IDE extension}
Besides anticipating a better user experience, we designed the prototype as an in-IDE extension
to more easily provide the code context to the LLM -- participants could select code to use as part of the context for a query.

\mysec{Pre-generated prompts}
We designed buttons that query the LLM with pre-generated prompts (\textit{prompt-less} interaction) to ask about an API, conceptual explanations, or usage examples, as shown in Figure~\ref{fig:tool}-\circled{4}.
We chose these based on API learning theory~\cite{Thayer.2021, Meng.2017, latoza2007program, duala2010information}, expecting this may particularly assist novice programmers or those unfamiliar with the APIs/domains or the LLM, as writing efficient search queries or prompts can be difficult for novices~\cite{Denny.2023, Ko.2011, Duala-ekoko.2012}.
At the same time, we also expected that this could reduce the cognitive burden of users in general in formulating prompts.

For {\small\texttt{Overview}} and the buttons {\small \texttt{API}, \texttt{Concept}, \texttt{Usage}}, we came up with prompt templates after a few iterations. 
To more efficiently provide the context to LLM, we used the library names and the list of API methods included in the selected code, such as ``Please provide a [library name] code example, mainly showing the usage of the following API calls: [list of API methods]'' for {\small\texttt{Usage}}.

\mysec{Unrestricted textual queries}
Users can also directly prompt the LLM (Figure~\ref{fig:tool}-\circled{5}), in which case \tool will automatically add any selected code as context for the query. Internally, the tool adds the selected code as part of the user prompt using pre-defined templates, and requests the LLM to respond based on the code context.

\mysec{Need-based explanation generation}
The tool is pull-based, i.e., it generates an explanation only when a user requests it. Similar to many previous developer information support tools, we wanted to reduce information overload and distraction. We expect that if and when enough context can be extracted from the IDE, hybrid (pull + push) tools will be possible, but this would require more research.

\mysec{Iterative design updates}
We ran design pilot studies and updated our prototype accordingly.
For example, we made the code summary as the default action for the tool trigger with code selection, after seeing pilot participants struggling to find parts of code to work on due to their unfamiliarity with libraries and domains. 
We updated the prompt-based interaction with LLM to support a conversational interface, based on the pilot participants' feedback that they wanted to probe the model based on their previous queries to clarify their intent or ask for further details. 
Finally, we opted to use GPT-3.5-turbo instead of GPT-4 as planned, after discovering that the response time was too slow in the pilot studies.

\section{Human Study Design}
\mysec{Participants}
We advertised our IRB-approved study widely within the university community (through Slack channels, posted flyers, and personal contacts) and to the public (through Twitter, email, and other channels). We asked each participant about their programming experience and screened out those who reported having a ``not at all'' experience. We did not ask about their professional programming experience, as the target users of our information support tools are not limited to professional developers. To minimize the possibility of participants knowing solutions, we specifically sought out participants who had not used the libraries included in our study. We accepted participants into the study on a rolling basis to capture a range of programming experience and existing domain knowledge. We compensated each participant with a \$25 Amazon Gift card. 
\revision{We recruited 33 participants and conducted 33 studies in total.
However, we had to exclude data from one participant from the analysis because they did not follow the instructions for using the extension. 
In the end, we had 9 women and 23 men participants.
Among them, 16 participants identified themselves primarily as students, 1 as software engineer, 2 as data scientists, and 13 as researchers. In the analysis, we divided the participants into two groups (students \vs professionals) based on this.
24 participants had experience with ChatGPT, 15 with Copilot, 5 with Bard, while 7 participants reported no prior use of any LLM-based developer tools.
In terms of familiarity with such tools, 14 participants stated that they have either used AI developer tools for their work or always use them for work.}

\begin{figure}
    \centering
    \includegraphics[width=\linewidth]{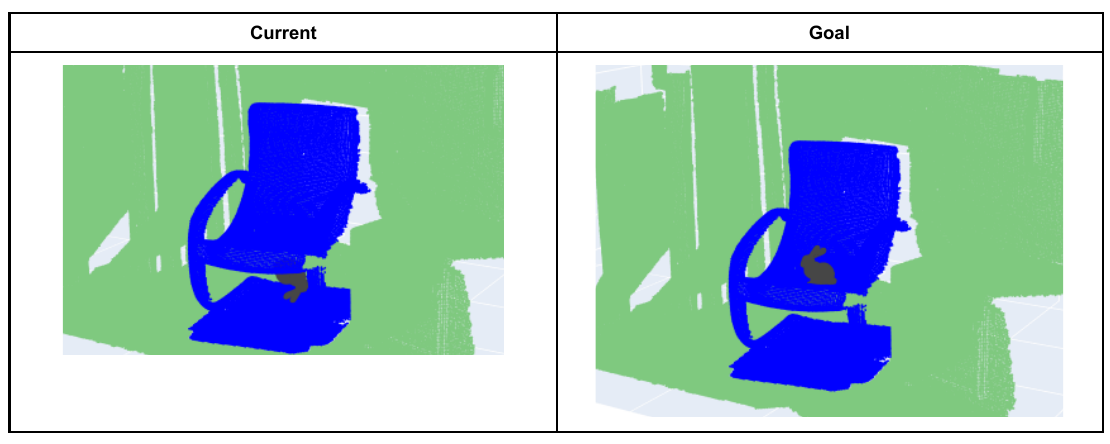}
    \caption{A 3D-rendering example sub-task (open3d-3). With these start and goal outputs, we asked the participants to ``Make the bunny sit upright on the chair.'' See Figure~\ref{fig:tool} for the corresponding starter code and the tool output. }
    \label{fig:task_eg}
\end{figure}

\mysec{Tasks}
The tasks were designed to simulate a scenario in which developers with specific requirements search the web or use existing LLMs to generate code and find similar code that does not precisely match their needs. 
For each task, we provided participants with a high-level goal of the code, start and goal outputs, and a starter code file loaded into the IDE. 
In this way, participants had to understand the starter code we provided and make changes to it so that the modified code met the goal requirements.
\revision{Each task consisted of 4 sub-tasks, to help reduce participant's overhead in planning, as well as to measure the task completion ratio in the analysis.
There were some subtle dependencies between the sub-tasks, so we advised participants to follow the order we gave, but they were free to skip.
Completing each sub-task required a small change, ranging from a single parameter value update to an addition of a line.
The difficulty levels of the sub-tasks varied, but we intentionally designed the first sub-task to be easy so that participants can onboard easily.
Sub-tasks also came with start and goal outputs and descriptions (see Figure~\ref{fig:task_eg}).} 
We did not include tasks that required strong programming knowledge, because our goal was 
to assess how well participants could understand the code.
For the same reason, we provided participants with starter code that was runnable and bug-free.

Our tasks cover both a common and less common domain that a Python developer might encounter in the wild. We chose two domains: data visualization and 3D rendering. These two tasks also allowed participants to easily check their progress, as they produce visual outputs that are comparable with the given goal.
For the data visualization task, we used the Bokeh~\cite{bokeh} library and asked participants to edit code that visualizes income and expenses data in polar plots. Understanding this code required knowledge of concepts related to visualizing data formats, marks, and data mapping. 
In the 3D rendering task, we used the Open3d~\cite{open3d} library. This task required knowledge of geometry and computer graphics. Participants were asked to edit code that involved point cloud rendering, transformation, and plane segmentation.

When selecting libraries, we intentionally did not choose the most common ones in their respective domains, to reduce the risk of participants knowing them well.
Choosing less common libraries also helped reduce the risk of an outsized advantage of our LLM-powered information generation tool. Responses for popular libraries can be significantly better than those for less commonly used ones, as the quality of LLM-generated answers depends on whether the LLM has seen enough relevant data during training.

The Bokeh starter code consisted of 101 LOC with 11 (6 unique) Bokeh API calls, and the starter code for the Open3D task consisted of 43 LOC with 18 (18 unique) Open3D API calls. The tasks were designed based on tutorial examples in each API's documentation.
In the starter codes, we did not include any comments in the code to isolate the effects of the information provided by our prototype or collected from search engines. All necessary API calls were in the starter code so participants did not need to find or add new ones.

In the task descriptions, we tried to avoid using domain-specific or library-specific keywords that could potentially provide participants with direct answers from either search engines or \tool. For instance, we opted to use ``make the bunny...'', instead of ``transform the bunny...'' which may have steered participants towards the function {\small\texttt{transform}} without much thought. 

The full task descriptions, starter code, solution for the demo, and the actual tasks are available in our replication package.

\mysec{Experimental Design}
We chose a within-subjects design, with participants using both \tool (treatment) and a search engine (control) for code understanding, but they did so on different tasks.
This allowed us to ask participants to rate both conditions and provide comparative feedback about both treatments.

The control-condition participants were not allowed to use our prototype, but they were free to use any search engine to find the information they needed.
The treatment-condition participants were encouraged to primarily use our prototype for information support. 
However, if they could not find a specific piece of information using our prototype, we allowed them to use search engines to find it. 
This was to prevent participants from being entirely blocked by the LLM. 
We expected that this was a realistic use case of any LLM-based tool, but it rarely happened during the study. 
Only 2 participants ended up using search engines during the treatment condition, but they could not complete the tasks even with the search engines.

We counterbalanced the tasks and the order they were presented to participants to prevent carryover effects, resulting in four groups (2 conditions x 2 orders). 
We used random block assignments when assigning participants to each group. 
Participants were assigned to each group to balance the self-reported programming and domain experience (data visualization and 3D rendering).
For every new participant, we randomly assigned them to a group that no previous participant with the same experience level had been assigned. 
If all groups had previous participants with the same experience level, we randomly assigned the participant to any of them.

\mysec{Study Protocol}
We conducted the study via a video conferencing tool and in person, with each session taking about 90 minutes; in-person participants also used the video conferencing tool, for consistency. 
At the beginning of the study, we asked participants to complete a pre-study survey, collecting their demographic information, background knowledge, and experience with LLMs.
We also estimated their general information processing and learning styles using a cognitive style survey~\cite{hamid2023measure} categorizing participants into two groups per dimension: comprehensive / selective information processing and process-oriented learning / tinkering. 
The participants were then asked to join our web-based VS Code IDE hosted on GitHub CodeSpaces~\cite{codespaces}, which provided a realistic IDE inside a web browser with edit, run, test, and debug capabilities, without requiring participants to install any software locally~\cite{davis2022s}. 
We then showed them a demo task description and explained what they would be working on during the real tasks. 
Before their first task in the treatment condition, we provided a tutorial for our plugin using the demo task, introducing each feature and giving three example prompts for the LLM.  
For the control condition, we did not provide any demo, as we expected every participant to be able to use search engines fluently. 
For each task, we gave participants 20 minutes to complete as many sub-tasks as they could.
During the task, we did not use the think-aloud protocol because we wanted to collect timing data. 
Instead, we collected qualitative data in the post-survey with open-ended questions. 
We also collected extensive event and interaction logs during the task.
After each task, we asked participants to complete a post-task survey to measure their understanding of the provided code and the API calls therein. 
At the very end, we asked them to complete a post-study survey where we asked them to evaluate the perceived usefulness and perceived ease of use of each code understanding approach and each feature in \tool. 
We based our questionnaire on the Technology Acceptance Model (TAM)~\cite{lee2003technology}, NASA Task Load Index (TLX)~\cite{hart1988development}, and pre- and post-study questionnaires that were previously used in similar studies~\cite{Ross.2023, Thayer.2021}.
See our replication package for the instruments. 

We conducted 33 studies in total, with 33 participants. The initial 18 studies were conducted on a one-on-one basis, while some studies in the latter half (involving 15 participants) were carried out in group sessions, with two to five participants simultaneously and one author serving as the moderator. We took great care to ensure that participants did not interrupt each other or share their progress. 
As mentioned before, we excluded one participant's data and used 32 participants' for the analysis.
We discovered this issue after the study, as this participant was part of the largest group session (with five participants).

\section{RQ1: Effects of \tool}
In this section, we report on the effectiveness of using \tool in understanding unfamiliar code.

\subsection{Data Collection}

\mysec{Code understanding}
To evaluate the effectiveness of each condition, we used three measurements:
(1) Task completion time: to complete each sub-task; (2) Task progress: we rated the correctness of the participants' solution to each sub-task and measured how many sub-tasks they correctly implemented; and (3) Understanding level: we cross-checked participants' general understanding of the starter code by giving them sets of quiz questions about the APIs included in the starter code. Each set contained three questions, requiring an in-depth understanding of the functionalities of each API call and the application domains.
To measure the effect of using \tool and search engines, we excluded the sub-tasks data if participants \textit{guessed} the solution without using the tool (\ie zero interaction with the tool) or search engines before completing it (\ie no search queries).

\mysec{Prior knowledge}
To control for prior knowledge, we used self-reported measures of participants' programming and domain experience. We expected more programming experience, especially in the specific domain, to lead to a faster understanding of code.

\mysec{Experience in AI developer tools}
Crafting effective prompts for LLM-based tools requires trial and error, even for NLP experts~\cite{Döderlein.2022, Denny.2023, zamfirescu2023johnny}. Therefore, we asked participants about their experience with LLM-based developer tools.
We expected participants' familiarity with other AI tools to affect their usage of the LLM-based information support tool, especially the use of free-form queries, and lead to more effective use of the extension than participants without such experience.

\subsection{Methodology}

To answer RQ1, we compared the effectiveness of using a \tool with traditional search engines for completing programming tasks by estimating regression models for three outcome variables. 
For task progress and code understanding, we used quasi-Poisson models because we are modeling count variables, and for the task completion time, we used a linear regression model.

To account for potential confounding factors, we included task experience, programming experience, and LLM knowledge as control variables in our models. 
Finally, we used a dummy variable ({\small\texttt{uses\_\tool}}) to indicate the condition (using \tool vs.\ using search engines). 
We considered mixed-effects regression but used fixed effects only, since each participant and task appear only once in the two conditions (with and without \tool).
For example, for the task completion time response, we estimate the model:

\noindent{\small\texttt{completion\_time $\sim$ domain\_experience + programming\_experience}}

\hfill {\small\texttt{ + AI\_tool\_familiarity + uses\_\tool
}}

\noindent The estimated coefficient for the {\small\texttt{uses\_\tool}} variable indicates the effect of using \tool while holding fixed the effects of programming experience, domain experience, and LLM knowledge.

\sisetup{
  input-symbols         = {()},
  group-digits          = false,
  table-space-text-post = ***,
  explicit-sign
}

\begin{table}[t]
\setlength{\tabcolsep}{4pt} 
    \centering
    \caption{\revision{Summaries of regressions estimating the effect of using the prototype. Each column summarizes the model for a different outcome variable. We report the coefficient estimates with the standard errors in parentheses.}}
    \label{tab:model}
    \sisetup{parse-numbers=false}
    \begin{tabular}{L{1.4cm}
    S[table-format=-3.2]
    S[table-format=-3.2]
    S[table-format=-3.2]
    S[table-format=-3.2]
    S[table-format=-3.2]}
    \toprule
                    & Progress       & Time (s)            & Underst$.$ & \multicolumn{2}{c}{Progress}      \\ 
                    & (1)            & (2)             & (3)               & Pros      & Students         \\ \midrule
    Constant        & 0.41      & 312.65    & -1.81**   & -0.38     & 1.82**    \\ 
                    & (0.49)    & (185.33)  & (0.89)    & (0.68)    & (0.83)    \\ [3pt]
    Domain          & 0.13*     & 23.14     & 0.41***   & 0.16      & 0.04      \\
    experience      & (0.07)    & (25.40)   & (0.12)    & (0.09)    & (0.11)    \\ [3pt]
    Program.        & -0.10     & -23.67    & 0.20      & 0.01      & -0.37*    \\
    experience      & (0.12)    & (43.53)   & (0.22)    & (0.17)    & (0.21)    \\ [3pt]
    AI tool         & -0.01     & 7.70      & -0.09     & 0.07      & -0.10     \\
    familiarity     & (0.07)    & (27.04)   & (0.14)    & (0.11)    & (0.10)    \\ [3pt]
    Uses \tool       &  0.47***  & -9.10     & 0.29      & 0.57**    & 0.29      \\ 
                    & (0.16)    & (57.26)   & (0.28)    & (0.22)    & (0.25)    \\ \midrule
    \makecell[tl]{$R^{2}$}  & 0.173  & 0.022  &   0.202 &   0.341  & 0.137 \\
    \makecell[tl]{Adj. $R^{2}$} & 0.117  & -0.046  &  0.148  &  0.243 &  0.010 \\ \bottomrule
    \multicolumn{6}{r}{\footnotesize Note: *p \textless 0.1; **p \textless 0.05; ***p \textless 0.01.}
    \end{tabular}
\end{table}

\subsection{Results}
Table~\ref{tab:model} columns (1)-(3) display the regression results for three response variables.
The task progress model (Table~\ref{tab:model}-(1)) shows a significant difference between the two conditions, with participants in the \tool condition completing statistically significantly more sub-tasks (0.47 more, $p$ < 0.01) than those who used search engines, controlling for experience levels and AI tool familiarity.
This indicates that \tool may assist users in making more progress in their tasks compared to search engines.

On the other hand, models (2) and (3) fail to show any significant difference in completion time and code understanding quiz scores between conditions.
This suggests that users in the \tool condition do not complete their tasks at a sufficiently different speed or have a sufficiently different level of understanding than those in the control group, given the statistical power of our experiment.

In summary, the results suggest that \tool may help users make more progress in their tasks without changing, for better or worse, their speed and code understanding abilities.

\subsection{Additional Analysis}
\label{ssec:gap}
After observing the significant effect of \tool on task progress, we dove deeper to examine whether all participants benefited equally from the tool. 
To do this, we divided the participants into two distinct groups based on their self-reported occupations (professionals and students) and estimated the effects of \tool usage in each group.\footnote{We considered but decided against, modeling interaction effects as they would have required more statistical power.} 
We opted for these groups as we did not have any prior theoretical framework to guide our grouping choices, and it provided a simple yet effective approach to group participants with multiple dimensions, including programming experience, skills, and attitude toward programming.


Although both groups were more successful when using the tool, there were notable differences in their performance gains. 
To better understand these variations, we estimated coefficients for each group (Table~\ref{tab:model}-Pros and -Students) and observed that the impact of \tool was significant only in the Pros group model. 
Specifically, professionals completed 0.57 more sub-tasks with \tool support compared to when they used search engines, whereas students did not experience significant gains.
These findings suggest that the degree of benefit provided by \tool may vary depending on participants' backgrounds or skills. 


\begin{summary}[RQ1]
There are statistically significant gains in task completion rate when using GILT, compared to a web search, but the degree of the benefit varies between students and professionals.
\end{summary}

\begin{figure}
    \centering
    \includegraphics[width=0.9\linewidth]{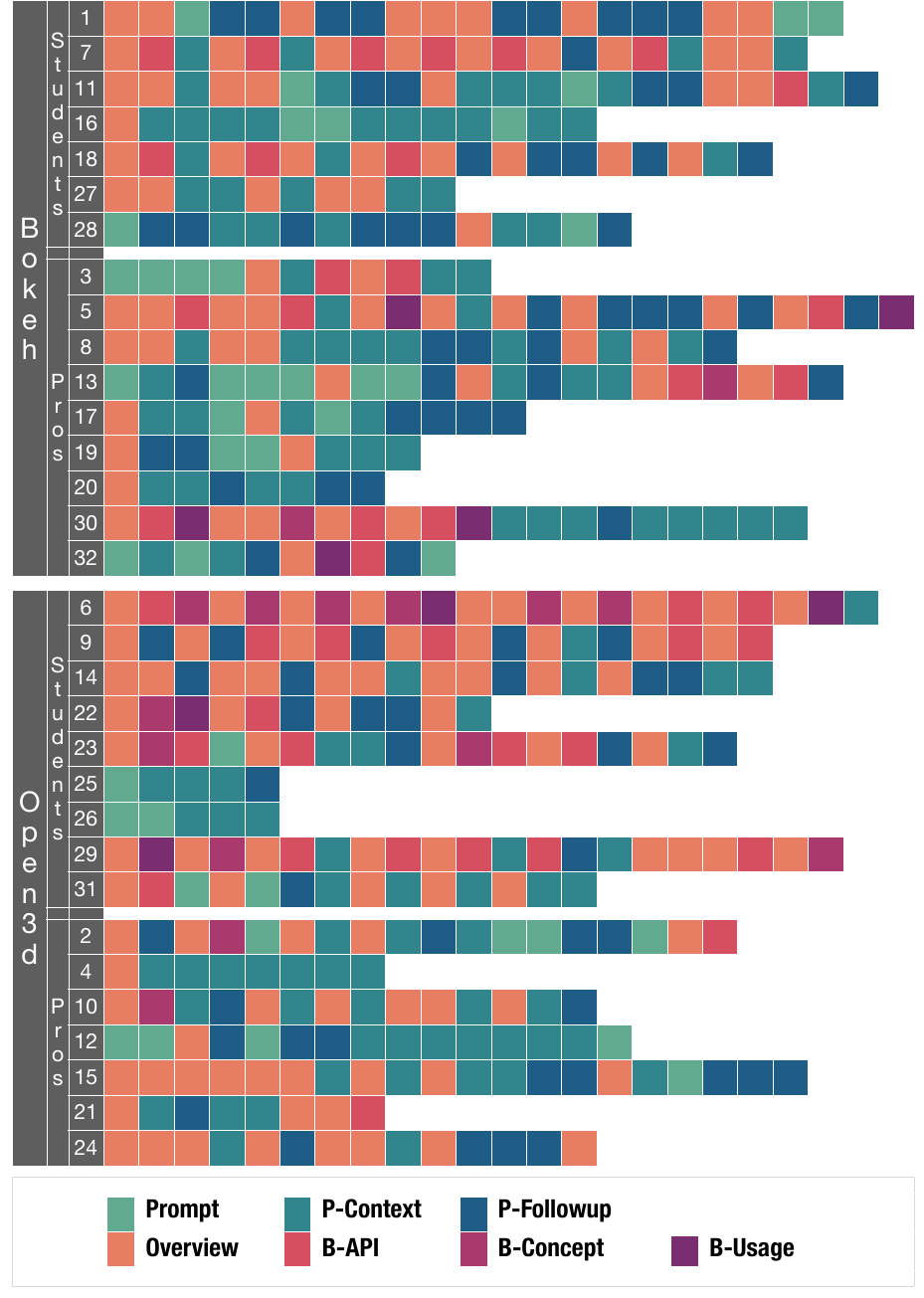}
    \caption{The sequences of feature usage in \tool. Each row corresponds to an individual participant, and the color cells are arranged chronologically, from left to right.}
    \label{fig:tool_usage}
    \vspace{-0.5cm}
\end{figure}

\section{RQ2: \tool Usage}

In this section, we focus on how participants interacted with \tool, their perception of the importance of different features, and how different factors correlate with the feature usage.


\subsection{Usage of Features} 

\revision{To analyze in more detail how participants actually used the tool, we instrumented \tool and recorded participants' event and interaction logs. The logs allowed us to count the number of times participants triggered each feature, and in what order.
To supplement the usage data, participants were asked to rate the importance of each feature in a post-task survey. We used these ratings to triangulate our findings from the usage data.}

\revision{
Figure~\ref{fig:tool_usage} summarizes the sequences of \tool features used by participants in the treatment condition.
On average, to complete their tasks in this condition, participants interacted with the LLM via \tool 15.34 times. The number of interactions per participant ranged from a minimum of 5 to a maximum of 23.}
The {\small\texttt{Overview}} feature was the most frequently used method to interact with the LLM, with an average of 4.76 activations per participant. Many participants also used {\small\texttt{Overview}} as their first feature, possibly because it requires minimal effort, with just a single click, in contrast to other features that necessitated the formulation of queries by participants, and perhaps also because some of the buttons (\eg ~{\small\texttt{Concept}}) required first using the {\small\texttt{Overview}} feature. Participants also frequently used {\small\texttt{Prompt-context}} (4.12 times) and {\small\texttt{Prompt-followup}} (2.88 times). General prompting without code context was used less frequently (1.27 times). While participants generally used buttons less frequently, some used them more frequently than queries (\eg P29), indicating personal preferences in prompt-based and prompt-less interactions. Specifically, the {\small\texttt{API}} button was used 1.24 times, the {\small\texttt{Concept}} button 0.45 times, and the {\small\texttt{Usage}} button 0.24 times on average.

\begin{figure}
    \centering
    \includegraphics[width=\linewidth]{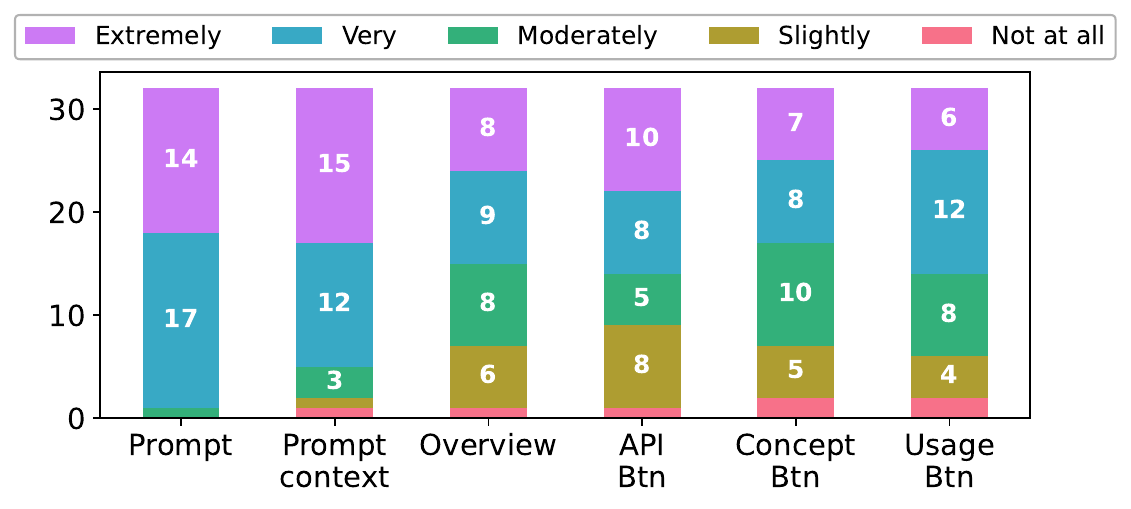}
    \caption{Participants' report on the importance of \tool features.}
    \label{fig:feature_importance}
    \vspace{-0.2cm}
\end{figure}

\begin{figure*}
\begin{minipage}[tl]{.3\textwidth}
    \centering
    \vspace{0pt}
    \includegraphics[width=\textwidth]{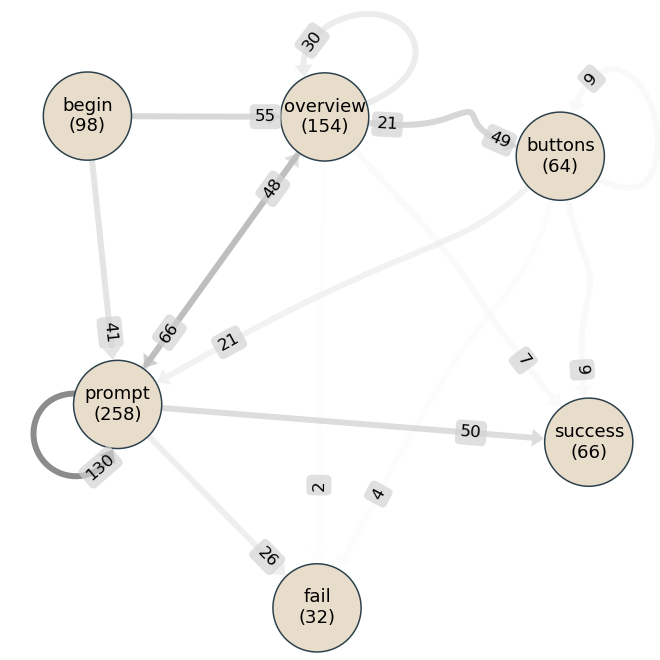}
    \\(a) All
\end{minipage}
\begin{minipage}[tl]{.3\textwidth}
    \centering
    \vspace{0pt}
    \includegraphics[width=\textwidth]{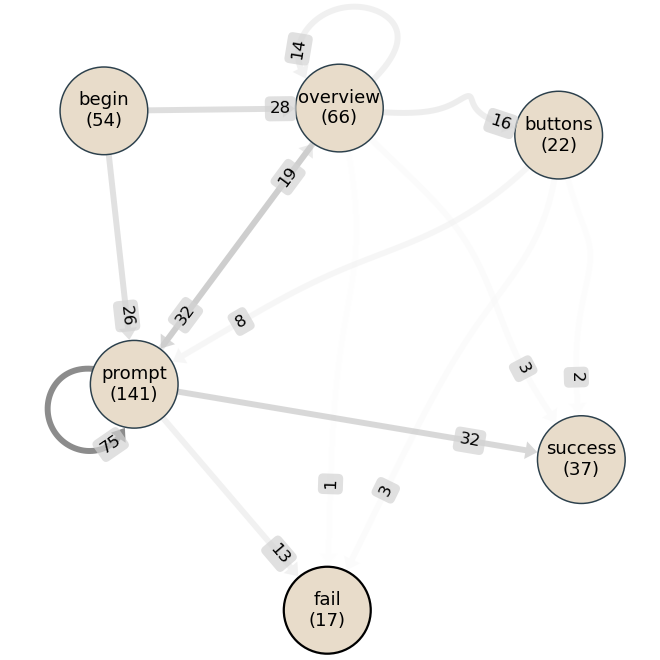}
    \\(b) Professionals
\end{minipage}
\begin{minipage}[tr]{.3\textwidth}
    \centering
    \vspace{0pt}
    \includegraphics[width=\textwidth]{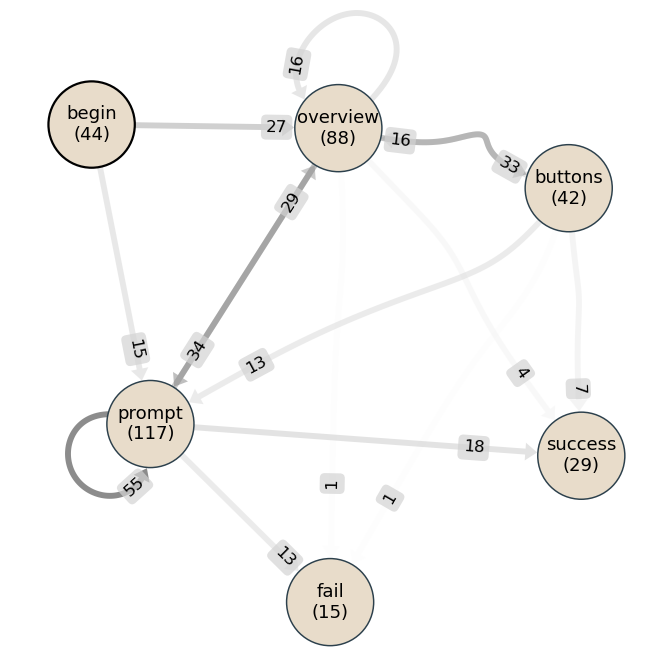}
    \\(c) Students
\end{minipage}
\caption{Transition Graphs for User Interaction. 
Each node displays the number of times users interacted with respective features, and each edge indicates the counted number of transitions between the connected features. For space and readability reasons, {\small\texttt{Prompt}}, {\small\texttt{Prompt-Context}}, and {\small\texttt{Prompt-Followup}} are merged into {\small\texttt{prompt}}, and {\small\texttt{API}}, {\small\texttt{Concept}}, and {\small\texttt{Usage}} are merged into {\small\texttt{buttons}}. Counts lower than 5 are omitted except for the edges connected to the `Success' and `Fail' nodes.}
\label{fig:transition}
\end{figure*}



The reported importance of the features by participants 
(see Figure~\ref{fig:feature_importance}) 
generally corresponds to the observed usage data. Most of the participants (97\%) responded that the ability to directly ask questions to the LLM was extremely/very important, whereas their reported usefulness of the buttons varied. The reported importance of the overview feature (53\% extremely/very important) was relatively low compared to the actual use, suggesting that participants may not have used the summary description provided by the overview but instead used it as context for further prompting or to activate buttons.

\revision{
To further investigate participants' interaction with the tool, we created transition graphs (Figure~\ref{fig:transition}) using sequences of feature use events for each sub-task, using both the sub-tasks successfully completed by participants and those that resulted in failure (due to incorrect answers or timeouts).
Out of the potential total of 128 sub-tasks (32 participants × 4 sub-tasks), 98 sub-tasks were started before the time ran out.
In understanding the transition graph, we focused on the last feature in each participant's sequence, with an assumption that when a participant completes a task, it is likely that the information from the last interaction satisfied their information needs.
Among the sub-tasks they successfully completed, a substantial majority (75\%) originated from prompt-based interactions. At the same time, 83\% of the failed tasks were also preceded by prompt-based interactions, so prompt-based interactions were not particularly likely to result in successful information seeking.}




\subsection{Professionals \vs Students}

\revision{
To better understand the experiences of professionals and students (see \Cref{ssec:gap}), we compared the transition graphs for both groups (Figure~\ref{fig:transition} (b) and (c)). 
Notable distinctions emerged in terms of the features more likely influencing the success and failure of sub-tasks. 
Specifically, for professionals, a majority (86\%) of successful sub-tasks originated from {\small\texttt{prompt}}, whereas for students, this percentage (62\%) was statistically significantly lower ($\chi^2$(1, 66), $p<.05$).
The success rate of prompt-based interaction was also higher among professionals (71\%: 32 out of 45) compared to students (58\%: 18 out of 31).
Conversely, the success rate of the overview and buttons for professionals (56\%: 5 out of 9) was lower than that of students (85\%: 11 out of 13).
These results may indicate that students, possibly with less experience in information seeking for programming, encounter challenges in formulating effective prompts compared to the professionals, and rely more on prompt-less interaction.
However, we can also infer that prompt-less interaction is still not sufficient to compete with the benefits of prompt-based interaction with the current design, as they only accounted for less than 40\% of the completed tasks.
}

\begin{table}[]
\centering
\caption{\revision{Frequencies of n-grams used differently in prompts by professionals and students. For clarity, we only include n-grams used uniquely by one of the two groups, with a frequency difference of more than 2. If multiple n-grams share the same longer n-gram, we report only the superset.}}
\label{tab:ngram}
\begin{tabular}{lL{5cm}ll}
\toprule
Sub-task & n-gram                                       & Pro.           & Stu. \\ \midrule
bokeh-2  & (`align', `text')                            & 3             & 0        \\
         & (`flip', `label')                            & 0             & 3        \\ 
bokeh-3  & (`annular', `wedge')                         & 6             & 0        \\
         & (`grid', `annular', `wedge')                 & 3             & 0        \\
         & (`first', `pie')                             & 3             & 0        \\
         & (`pie', `chart')                             & 3             & 0        \\
         & (`add', `legend')                            & 0             & 4        \\
         & (`tell', `line', `need', `change')           & 0             & 3        \\
o3d-3    & (`sit', `upright', `chair')                  & 4             & 0        \\
         & (`make', `bunny', `sit', `upright', `chair') & 3             & 0        \\ \bottomrule
\end{tabular}
\end{table}

\revision{
To further investigate the differences in the two groups' prompt engineering, we analyzed the text of the prompts they wrote, by comparing the frequencies of bi-, tri-, and quadrigrams in the prompts.
Table~\ref{tab:ngram} presents the list of n-grams that showed divergent usage between the two groups. 
One notable observation is that the n-grams used by the professional group include more effective keywords, or they revise the prompts to incorporate such keywords. 
For instance, in the bokeh-3 sub-task, none of the participants in the student group used the critical keyword ``annular wedge,'' which is essential for generating the information needed to solve the task, although it was used multiple times in the provided starting code. 
Instead, students tended to use more general keywords or keywords that had a different concept in the library (\eg ``legend'') and faced difficulties in effectively revising the prompts.
In addition, more participants in the professional group demonstrated proficiency in refining their prompts by providing further specifications.
For example, one participant revised the prompt from ``How to change the position of the bunny to 180 degrees'' to ``How to transform the bunny\_mesh to 180 degrees.''
We infer that the difference in the benefit received from \tool by the two groups can be at least partially attributed to their proficiency in prompt engineering.
}


\subsection{Other Factors Associated with Feature Use}

During the pilot studies, we observed that participants approached the tasks differently depending on their familiarity with other LLM-based tools, and styles of information processing and learning as observed in many previous studies on software documentation and debugging~\cite{Meng.2017, Meng.2018, beckwith2006tinkering}.
Thus, we tested whether the \tool feature use correlates with factors other than their experience.

\mysec{Hypotheses}
Out of two information processing styles~\cite{darley1995gender, burnett2016gendermag, grigoreanu2010strategy}, people who exhibit a ``selective information processor'' tendency focus on the first promising option, pursuing it deeply before seeking additional information. 
On the other hand, people who are ``comprehensive information processors'' tend to gather information broadly to form a complete understanding of the problem before attempting to solve it. Based on these processing styles, we hypothesized that selective processors would utilize \tool's {\small\texttt{Prompt-followup}}, as they would prefer to use a depth-first strategy.

In terms of learning styles~\cite{perkins1986conditions, burnett2016gendermag}, ``process-oriented learners'' prefer tutorials and how-to videos, while ``tinkerers'' like to experiment with features to develop their own understanding of the software's inner workings. Consequently, we hypothesized that tinkerers would use \tool less often, as they would prefer to tinker with the source code rather than collect information from the tool.

\begin{table}
\centering
\caption{\revision{Summaries of regressions testing for associations between the user factors and the feature usage counts. Each column summarizes a regression modeling a different outcome variables. We report the coefficient estimates with their standard errors in parentheses.}}
\label{tab:usage}
\begin{tabular}{lS[table-format=-3.2]S[table-format=-3.2]S[table-format=-3.2]}
\toprule
                                    & \multicolumn{1}{c}{Prompt}        & \multicolumn{1}{c}{Followup}           & \multicolumn{1}{c}{All}         \\ 
                                    & \multicolumn{1}{c}{(1)}            & \multicolumn{1}{c}{(2)}                & \multicolumn{1}{c}{(3)}         \\ \midrule
Constant                        & 1.39***       & -0.82         & 2.43***       \\
                                & (0.31)        & (0.69)        & (0.27)        \\ 
AI tool & 0.19**        & 0.38**        & 0.11          \\
familiarity & (0.07)        & (0.15)        & (0.06)        \\ [2pt]
Infomation  & -0.04         & 0.44          & -0.04         \\
Comprh.  & (1.15)        & (0.30)        & (0.13)        \\  [2pt]
Learning& 0.19          & 0.60**        & -0.12         \\
Process & (1.14)        & (0.29)        & (1.13)        \\  \midrule
\makecell[tl]{$R^{2}$}              & 0.262     \ \                  \    & 0.283                        \  \       &                    0.165 \ \ \    \\
\makecell[tl]{Adj. $R^{2}$}         & 0.184     \ \                  \    & 0.206                        \  \       &                    0.075 \ \ \    \\ \bottomrule
\multicolumn{4}{r}{\footnotesize Note: *p \textless 0.1; **p \textless 0.05; ***p \textless 0.01.}
\end{tabular}
\vspace{-1cm}
\end{table}

We also expected that participants who were already familiar with LLM-based tools would use prompt-based interaction in general ({\small\texttt{Prompt}}), especially the chat feature, more frequently, as they would already be accustomed to using chat interfaces to interact with LLMs. Conversely, we posited that participants with less experience with such tools would use the buttons more, as prompt engineering might be less familiar to them and place greater cognitive demands on them.

\mysec{Methodology}
To test for associations between \tool features used and the factors above we again used multiple regression analysis. 
We estimated three models, each focused on one particular feature. For each model, the dependent variable was the feature usage count, while participants' information processing style, learning style, and familiarity with AI developer tools were modeled as independent variables to explain the variation in usage counts.

\mysec{Results}
Table~\ref{tab:usage} presents the results of the regression analysis conducted for three response variables. 
The first model (Prompt (1)), which uses the total count of prompt-based interactions ({\small\texttt{prompt} + \texttt{prompt-context} + \texttt{prompt-followup}}), 
reveals that developers who are more familiar with other AI developer tools are more likely to prompt the LLMs using natural language queries. This result confirms our hypothesis that the AI tool familiarity level influences developers' use of queries.
The familiarity level also has a statistically significant impact on {\small\texttt{prompt-followup}}, as shown in the Followup model (2). However, we did not find any significant impact of participants' information processing style on their use of \tool. This means that selective processors and comprehensive processors probed the LLMs similarly, as far as we can tell. 
The model, however, shows a statistically significant correlation between participants' learning styles and {\small\texttt{prompt-followup}} feature usage. Specifically, process-oriented learners tend to probe LLMs more frequently than tinkerers. 
This result might indicate that process-oriented learners are more likely to learn thoroughly before proceeding to the next step, while tinkerers tend to tinker with the code after getting the minimum amount of direction from \tool.
Finally, the All model (3), which uses the total count of all \tool interactions, indicates that there is no statistically significant difference between the information styles, learning styles, and familiarity levels in terms of overall feature usage counts. 

\begin{summary}[RQ2]
Overall, participants used \texttt{Overview} and \texttt{Prompt-context} most frequently. However, the way participants interact with GILT varied based on their learning styles and familiarity with other AI tools. 
\end{summary}

\section{RQ3: User Perceptions}

In this section, we investigate how participants perceived their experience of using \tool. Specifically, we examine their perceived usefulness, usability, and cognitive load in comparison to search-based information seeking. Additionally, we explore the pros and cons participants reported, and suggestions for improving the tool.


\subsection{Comparison with Web Search}

We employed two wildly-used standard measures, TLX and TAM, in our post-task survey and compared them using two-tailed paired t-tests.
TAM (Technology Acceptance Model)~\cite{lee2003technology} is a widely used survey that assesses users' acceptance and adoption of new technologies, and TLX (Task Load Index)~\cite{hart1988development} is a subjective measure of mental workload that considers several dimensions, including mental, physical, and temporal demand, effort, frustration, and performance.
\revision{The summaries of TAM and TLX comparisons can be found in our replication package.}

The average scores for the [perceived usefulness, perceived ease of use] in TAM scales were [27.3, 29.75] for the control condition, and [33.49, 34.2] for the treatment condition.
The paired t-tests on the TAM scores indicated that there were significant differences in perceived usefulness and perceived usability scores between the two conditions ($p < 0.001$). Specifically, participants rated \tool higher on both dimensions than they did search engines.

For TLX items [mental demand, physical demand, temporal demand, performance, effort, frustration], the average scores were [3.8, -2.1, 4.0, 1.6, 3.4, -0.1] for the control condition and [3.3, -2.5, 2.6, 3.3, 3.3, 1.0] for the treatment condition. Paired t-tests on the TLX scores revealed statistically significant differences between the tool and search engines in temporal demand ($p < 0.05$) and performance ($p < 0.05$) but not in other items. These results indicate that the participants felt less rushed when using \tool than when using search engines, and they felt more successful in accomplishing the task with the tool than with search engines, but there were no significant differences in other dimensions.


\subsection{User Feedback}
In the post-task survey, we asked open-ended questions regarding (i) their general experience with using \tool, (ii) the tool's fit with the participants' workflow, (iii) comparison with other tools, and (iv) opportunities for improvement.
Two authors conducted a thematic analysis~\cite{clarke2013teaching} to analyze the answers.
Initially, two authors separately performed open coding on the same set of 8 responses (25\% of the entire data), and convened to discuss and merge the codes into a shared codebook. The first author coded the rest of the responses and discussed with the rest of the authors whenever new codes needed to be added.
The codebook is available in our replication package, and we discuss some of them here.

\revision{The participants in this study reported several positive aspects of the tool, with the most notable being context incorporation.
Participants valued the ability to prompt the LLM with their code as context, which allowed them to tailor the LLM's suggestions to their specific programming context, \eg
``the extension generated code that could easily be used in the context of the task I was performing, without much modification.'' (P5)
Participants also found it extremely useful to prompt the LLM with just code context, as it allowed them to bypass the need to write proficient queries, a well-known challenge in search-based information seeking~\cite{Ko.2011, xia2017developers}. 
P15 mentioned 
``\textit{It's nice not to need to know anything about the context before being effective in your search strategy.}''
}

\revision{
Many participants reported that using the tool helped them speed up their information seeking, by reducing the need to forage for information, \eg ``Stack Overflow or a Google search would require more time and effort in order to find the exact issue and hence would be time-consuming.'' (P27)
}

Some participants, however, reported having a hard time finding a good prompt that could give them the desired response. Combined with the need for good prompts and the limitations of LLM, this led some participants to report that the responses provided by the tool were occasionally inaccurate, reducing their productivity. P28 summarized this issue well:
``\textit{[prototype] was not able to give me the code that I was looking for, so it took up all my time (which I got very annoyed about). I think I just didn't word the question well.}''

Participants had mixed opinions on the different features of the tool, especially the buttons. Some preferred to use ``\textit{different buttons for different types of information so I didn't have to read a lot of text to find what I was looking for}'' (P7), while others thought that was overkill and mentioned ``\textit{a simpler view would be nice.}'' (P8)

Compared to ChatGPT, 17 participants (out of 19 who answered) mentioned advantages of \tool, with the {\small\texttt{Prompt-} \texttt{context}} feature being one of the main ones.
Participants expressed positive feelings about CoPilot but acknowledged that the tool had a different role than CoPilot and that they would be complementary to each other,  e.g.: 
``\textit{Copilot is a tool that I can complete mechanical works quickly, but [\tool] offers insight into more challenging tasks.}'' (P29)

Many participants reported that the tool would be even more useful when combined with search engines, API documentation, or CoPilot,\footnote{Notably, GitHub independently announced these enhancements to Copilot already, after we conducted our study: \url{https://www.theverge.com/2023/7/20/23801498/github-copilot-x-chat-code-chatbot-public-beta}} as they provide different types of information than the tool. Having the ability to choose sources based on their needs would enhance their productivity by giving them control over the trade-offs, such as speed, correctness, and adaptability of the information.

\begin{summary}[RQ3]
Participants appreciated GILT's ability to easily incorporate their code as context, but some participants reported that writing a good prompt is still a challenge. 
\end{summary}

\section{Threats to Validity}
One potential concern with our study design is the task and library selection. 
We only used tasks that show visible outputs, which might have led participants to detect potential errors more easily, compared to other tasks, such as optimization or parallel programming. 
However, we believe that the tasks we chose are representative of common programming errors that would need to be identified in real-world programming situations. Indeed, when we asked the participants in the post-task survey, both data visualization and 3D rendering tasks were reported to very or extremely closely resemble real-world tasks by 82\% and 73\% of the participants.

Similarly, the selection of libraries might have biased the study results. 
However, in selecting libraries for our study, we avoided using popular libraries that could unintentionally give an advantage to LLMs. 
We believe that the libraries we chose are of medium size and quality, and therefore represent a fair test of the LLM tools. 
However, it is possible that different libraries or larger codebases could produce different results.

Despite our efforts to create a controlled experience, several factors differentiate our in-IDE extension from search engines, aside from the inclusion of LLMs. 
For example, although previous research investigating the incorporation of search into IDE did not find a statistically significant difference between the control and treatment groups~\cite{li2015amassist, brandt2010example}, the in-IDE design itself may have been more helpful than access to LLMs, as it potentially reduced context-switching. 
Thus, further studies are needed to gain a better understanding of the extent to which each benefit of our prototype can be attributed to these differences.

Additionally, the laboratory setting may not fully capture the complexity of real-world programming tasks, which could impact the generalizability of our findings. 
Also, the time pressure participants could have felt, and the novelty effect in a lab setting could have changed how users interact with LLMs.
Our sample size, 32, was relatively small and skewed towards those in academia. 
This may also limit the generalizability of our findings to more professional programmers. 
Thus, future research with larger, more diverse samples is necessary to confirm and expand upon our results.

Our analysis also has the standard threats to statistical conclusion validity affecting regression models.
Overall, we took several steps to increase the robustness of our estimated regression results. First, we removed outliers from the top 1\% most extreme values.  Second, we checked for multicollinearity using the Variation Influence Factor (VIF) and confirmed that all variables we used had VIF lower than 2.5 following \citet{johnston2018confounding}.

Another potential threat to the validity of our findings is the rapid pace of technological development in the field of LLM tools. 
Despite our efforts to use the most up-to-date LLM available at the time of the submission, it is possible that new breakthroughs could render our findings obsolete before long. 

\section{Discussion and Implications}

\mysec{Comprehension outsourcing}
Our analysis revealed an intriguing finding regarding participants' behavior during the study, where some of them deferred their need for code comprehension to the LLM, which was well described by one participant as \textit{comprehension outsourcing}.
These participants prompted the model at a higher level directly and did not read and fully comprehend the code before making changes.
As one participant commented, ``\textit{I was surprised by how little I had to know about (or even read) the starter code before I can jump in and make changes.}''
This behavior might be attributed to developers' inclination to focus on task completion rather than comprehending the software, as reported in the literature~\cite{Maalej.2014}. 
Or, participants may have also weighed the costs and risks of comprehending code themselves, and chosen to defer their comprehension efforts to the language model.
While this behavior was observed in the controlled setting of a lab study and may not fully reflect how developers approach code comprehension in their daily work, it does raise concerns about the potential impact of such a trend (or over-reliance on LLMs~\cite{Vaithilingam.2022}) on code quality. 
This highlights the importance of preventing developers who tend to defer their comprehension efforts to the LLM from being steered in directions that neither they nor the LLM are adequately equipped to handle. 
Studies showing developers' heavy reliance on Stack Overflow, despite its known limitations in accuracy and currency~\cite{zhang2018code, wu2019developers}, further emphasize the need for caution before widely adopting LLM-based tools in code development.
Research on developers' motivations and reasons for code comprehension when LLMs are available will be valuable in informing future tool designs.


\mysec{Need for more research in UI}
\revision{In our analysis, we observed a notable trend where the professionals benefited \emph{more} from the tool compared to students.
Our examination of the prompts indicated that this discrepancy may arise because students face challenges in constructing effective queries or revising them to obtain useful information, aligning with findings in the literature on code generation using LLMs~\cite{Denny.2023}. 
Although we provided an option to use prompt-less interaction with LLMs to reduce the difficulty in prompt engineering, a lot of participants chose to use prompt-based interaction, possibly due to their familiarity with other AI tools, the potentially higher quality of information this mode produces, or other reasons that our study did not cover.
However, we find our results still promising, as we observed that students used prompt-less interaction more than the professionals and succeeded more when using the buttons than using the prompts.
We believe that further research is needed, exploring various interaction options to support a diverse developer population. 
}

\mysec{Utilize more context}
\cameraready{One of the main advantages of \tool reported by the participants is its ability to prompt the LLM with the code being edited as context.
We believe that additional types of context can be leveraged to improve the tool's utility, including project context (e.g., project scale and domain), system context (e.g., programming languages and target deployment environments), and personal context (e.g., programming expertise in libraries, and domains). By combining these contexts with proper back-end engineering, we believe that \tool, or other LLM-powered developer tools, will be able to provide relevant information to developers with even less prompt engineering efforts of the users.}


\mysec{Need further studies in real-world settings}
One possible explanation for some of the models with null results from RQ1 and RQ2 is the artificial setting of the lab study, where participants were encouraged to focus on small, specific task requirements instead of exploring the broader information dimension. 
For example, participants prioritized completing more tasks rather than fully understanding the code, as reported by participant P18 in their survey response: ``\textit{ [\tool] ..., which could definitely help one to tackle the task better if there weren't under the timed-settings.}''
Thus, although our first study shed some light on the potential challenges and promises, to fully understand the implications of deploying this tool into general developer pipelines, it is necessary to observe how programmers use it in real-world settings with larger-scale software systems, less specific goals, and over a longer time frame.
Given that GitHub recently launched CopilotX~\cite{aaaaa}, a tool that offers a comparable set of features to our prototype to enhance developer experience, such research is urgently needed.
We believe that our findings are a timely contribution and a good first step for 
researchers and tool builders in designing and developing developer assistants that effectively use LLMs.

\section{Conclusion}
We presented the results of a user study that aimed to investigate the effectiveness of generation-based information support using LLMs to aid developers in code understanding. With our in-IDE prototype tool, \tool,
we demonstrated that this approach significantly enhances developers' ability to complete tasks compared to traditional search-based information seeking. 
At the same time, we also identified that the degree of benefits developers can get from the tool differs between students and professionals, and the way developers interact with the tool varies based on their learning styles and familiarity with other AI tools.

\mysec{Data Availability}
Our supplementary material includes the replication package, including the study protocol, tasks, study data, scripts to replicate the analyses, as well as the prototype tool, GILT, \cameraready{are available online at }\DOIbox{https://doi.org/10.5281/zenodo.10461385}~\cite{nam_2024_10461385}.

\section{Acknowledgement}
\cameraready{We would like to thank our participants for their time and input, and our reviewers for their valuable feedback.}

\bibliographystyle{ACM-Reference-Format}
\bibliography{reference}
\balance


\end{document}